\documentclass{aastex63}
\usepackage{amssymb}
\usepackage{amsmath}
\usepackage{braket}
\usepackage{mathrsfs}
\usepackage{mathtools}
\usepackage{array}

\usepackage{bm}
\usepackage{url}

\received{}
\revised{}
\accepted{}

\submitjournal{}

\shorttitle{Relation between the variances of distortions
of GWs owing to gravitational lensing}
\shortauthors{Makoto Inamori and Teruaki Suyama}

\renewcommand{\theequation}{\thesection.\arabic{equation}}

\newcommand{\f}{\mathrm{f}}

\newcommand{\N}{\mathcal{N}}

\newcommand{\x}{{\bm x}}
\newcommand{\y}{\bm y}
\newcommand{\n}{\mathrm{n}}
\newcommand{\m}{\mathrm{m}}
\newcommand{\I}{\mathrm{i}}

\newcommand{\D}{d}

\def\be{\begin{equation}}
\def\ee{\end{equation}}
\def\ben{\begin{eqnarray}}
\def\een{\end{eqnarray}}

\begin{document}

\title{
Universal relation between the variances of distortions
of gravitational waves owing to \\
gravitational lensing
}

\author{Makoto Inamori}
\author{Teruaki Suyama}

\affiliation{
Department of Physics, Tokyo Institute of Technology, 2-12-1 Ookayama, Meguro-ku,
Tokyo 152-8551, Japan
}

\begin{abstract}
Gravitational waves from the distant sources are 
gravitationally lensed
during their propagation through the intervening matter inhomogeneities 
before arriving at detectors.
It has been proposed in the literature that the variance of the 
lensed waveform
can be used to extract information of the matter power spectrum at very small scales 
and of low-mass dark halos.
In this Letter, we show that the variance of the amplitude fluctuation and
that of the phase fluctuation of the lensed waveform obey a 
simple relation irrespective of the shape of the matter power spectrum.
We study conditions under which this relation can be violated
and discuss some potential applications of the relation.
This relation may be used to confirm the robustness
of claimed observations of gravitational lensing of gravitational waves
and the subsequent reconstruction of the matter power spectrum.
\end{abstract}

\renewcommand{\thepage}{\arabic{page}}
\setcounter{page}{1}
\renewcommand{\thefootnote}{\#\arabic{footnote}}
\setcounter{footnote}{0}

\section{Introduction}
Direct detections of gravitational waves (GWs) 
have opened the golden era of the GW astronomy \citep{Abbott:2020niy}.
It is highly expected that many new discoveries by the GW experiments will excite
us in the coming decades and progress our understanding about the universe (e.g. \citet{Bailes:2021tot}).
One of such discoveries is the gravitational lensing of the GWs \citep{Schneider}.
General relativity (GR) predicts that GWs propagating in the gravitational 
potential sourced by matter distribution are gravitationally lensed in a
similar manner to light \citep{MTW}.
Gravitational lensing typically magnifies gravitational-wave amplitude
and may significantly contribute
to the high mass tail of the mass distribution of the black hole binaries \citep{Dai:2016igl, Oguri:2018muv}.
One notable difference between the gravitational lensing of the GWs and that of light 
is that, due to the long-wavelength nature of the GWs from astrophysical sources,
wave effects such as diffraction become important 
in some cases for which wave optics
must be used, while the geometrical optics is an excellent approximation for light \citep{Ohanian:1974ys, Nakamura:1997sw, Nakamura:1999}\footnote{However, there is an exception.
For the very compact lensing objects such as primordial black holes,
the wave optics plays an important role even for light \citep{Gould:1992ApJ, Sugiyama:2019dgt}.}.

In the regime of wave optics, the amplification factor, which represents the
amount of distortion of the wave by the gravitational lensing,
depends on the wave frequency differently from what it does in the geometrical optics \citep{Nakamura:1999}. 
Because of this, the lensed waveform in the wave optics provides us with
additional information about the lensing objects that the geometrical optics does not.
For instance, even if the source position is outside the Einstein radius, 
for which there is only a single path in the geometrical optics, 
the mass of the lensing object may be extracted in the wave optics \citep{Takahashi:2003ix}.
Another important feature of the wave effects is that the waves are 
insensitive to structures smaller than the Fresnel scale 
$r_F$ which scales with the wave frequency $f$ as $\propto 1/\sqrt{f}$ \citep{Macquart:2004sh, Takahashi:2005ug}.
Thus, measurement of the amplification factor at different frequencies,
which is possible for GWs whose frequency varies in time as it is the case
for the GWs emitted from the chirping binaries, 
allows us to probe the matter distribution at different 
Fresnel scales 
\citep{Takahashi:2005ug, Oguri:2020ldf, Choi:2021jqn}.

In \citet{Takahashi:2005ug}, a novel idea was proposed that 
measurements of the variance of the amplitude and phase fluctuations of the amplification factor
enable us to determine the matter power spectrum at the Fresnel scale.
As an example, for the GWs in the decihertz range which corresponds 
to the sensitivity range of the future space interferometers such as DECIGO \citep{Seto:2001qf},
the Fresnel scale is about $1~{\rm pc}$ (see Eq.~(\ref{Fresnel-2})).
The matter inhomogeneities at this scale are supposed to be dominated by low-mass dark halos,
and information of the matter power spectrum around that scale will provide us with
the fundamental properties of dark matter and possibly new knowledge of the primordial
power spectrum.
In \citet{Oguri:2020ldf}, as an extension of the previous work \citep{Takahashi:2005ug}, 
detectability of the low-mass dark halos as well as the primordial black holes has
been investigated, and it is concluded that the measurements of the gravitational lensing
variance are a promising achievement in the future GW observations.
Because of the fundamental importance of the variance of the amplification factor of the GWs, there is a good motivation to study its basic properties.

In this Letter, we point out that there is an intriguing consistency relation
between the variance of the amplitude and the phase fluctuations of the amplification factor
that has not been given in the literature.
Remarkably, this relation holds true irrespective of the shape of the matter power spectrum.
Thus, this relation may provide a consistency test to evaluate if the observational determination 
of the variance of the gravitational
lensing has been done correctly and allow us to conduct the subsequent reconstruction of the matter power spectrum on a solid basis.
We also study in which situations the consistency relation can be violated
and discuss some potential applications of the consistency relation.

\section{Weak gravitational lensing of GWs}
\subsection{Amplification factor}
In this subsection, we briefly overview the formulation of the gravitational lensing
of GWs by the matter inhomogeneities. 
This overview is intended to make this Letter self-contained and hence the content is 
minimal.
Those who want to know more about individual equations and statements are recommended to read 
\citet{Oguri:2020ldf} and references therein.

\begin{figure}[t]
  \begin{center}
    \includegraphics[clip,width=10.0cm]{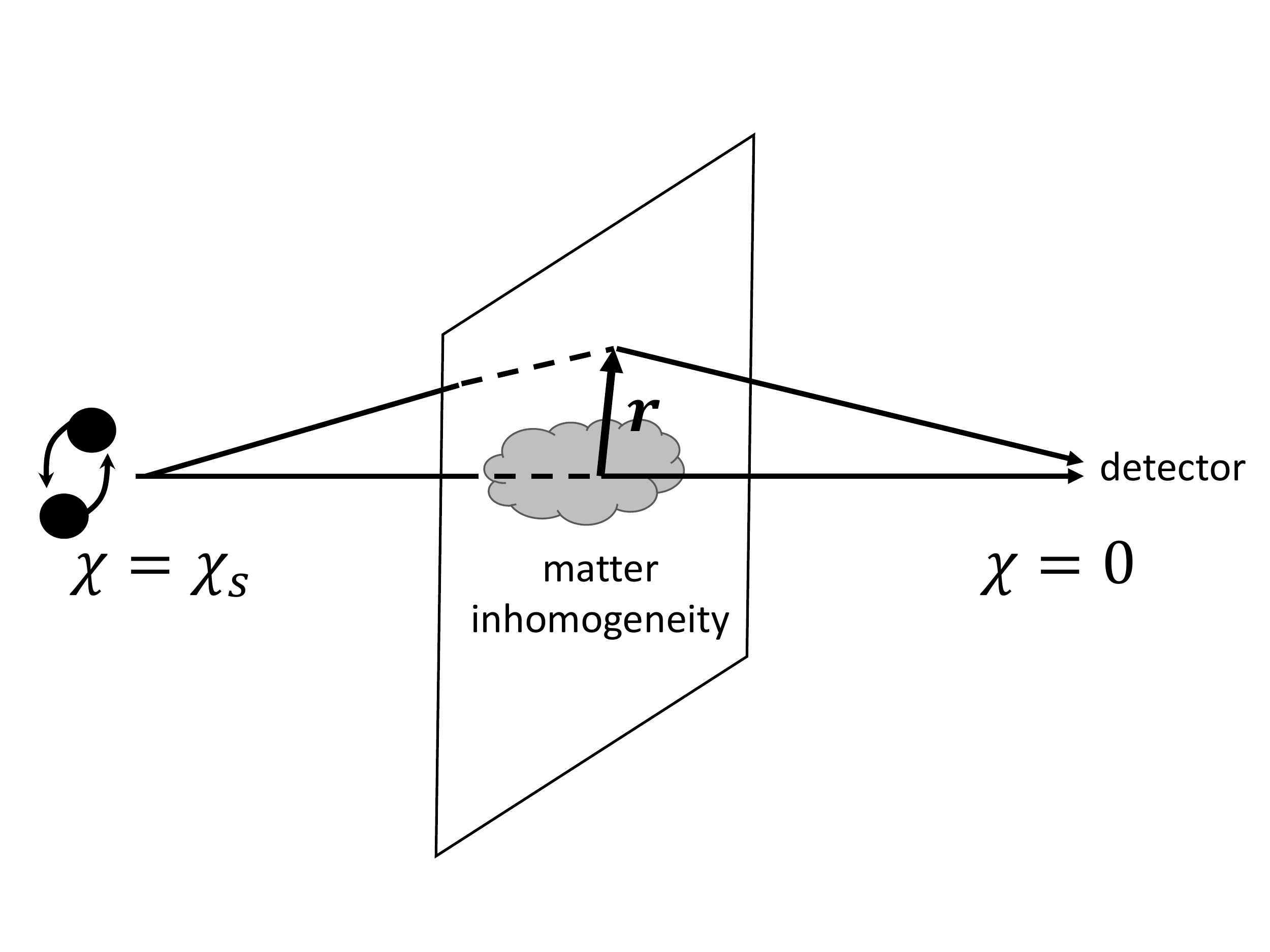}
    \caption{Schematic configuration of the gravitational lensing of the GWs
    by the intervening matter inhomogeneities.}
    \label{fig1}
  \end{center}
\end{figure}

Ignoring the tiny variation of the polarization of the GWs by the lensing objects, 
the amplitude $\phi$ of the lensed
gravitational waves at the detector's position 
is represented, in the frequency domain, 
by the product of the amplification factor $F$ and the unlensed wave $\phi_0$ as
\be
\phi (f)=F(f) \phi_0 (f).
\ee
Here $f$ is the (comoving) frequency of the GWs.
Throughout this Letter, we assume weak lensing for which the deviation of $F$
from unity is given by the linear order in the gravitational potential $\Phi$
sourced by the matter inhomogeneities (Born approximation).
In this regime, the amplification factor of the gravitational waves
emitted from the source at the comoving distance $\chi_s$ from the detector
is given by \citep{Takahashi:2005sxa}
\be
F(f)-1=-4\pi f^2 \int_0^{\chi_s} d\chi \frac{\chi_s}{\chi (\chi_s-\chi)}
\int d^2 r~ \Phi (\chi,{\bm r};t(\chi)) e^{2\pi if \Delta t({\bm r})}.
\label{born-app}
\ee
Here $\chi$ is the comoving distance from the detector,
${\bm r}$ is the two-dimensional vector perpendicular to the line of sight,
$t(\chi)$ is the cosmic time when the wave is at $\chi$,
and $\Delta t$ is the geometric time delay given by
\be
\label{geometrical-t}
\Delta t({\bm r})=\frac{\chi_s}{2\chi (\chi_s-\chi)} {\bm r}^2.
\ee
See Fig.~\ref{fig1} as a schematic picture representing the configuration.
For future convenience, we introduce the Fourier transformation of $\Phi$ as
\be
\Phi (\chi, {\bm r};t)=\int \frac{dk_\parallel}{2\pi}
\int \frac{d^2 k_\perp}{{(2\pi)}^2} ~{\tilde \Phi} (k_\parallel, {\bm k_\perp};t)
e^{ik_\parallel \chi+i {\bm k_\perp}\cdot {\bm r}}.
\ee
Substituting this expression into Eq.~(\ref{born-app}) and performing
integration over ${\bm r}$, we obtain
\be
F(f)-1=-4\pi i f\int_0^{\chi_s} d\chi
\int \frac{dk_\parallel}{2\pi}
\int \frac{d^2 k_\perp}{{(2\pi)}^2} ~{\tilde \Phi} (k_\parallel, {\bm k_\perp};t(\chi))
e^{ik_\parallel \chi-i \frac{\chi (\chi_s-\chi)}{4\pi f \chi_s}{\bm k_\perp}^2}.
\ee

The amplification factor is a complex number.
In physical terms, the absolute value and the argument of complex of 
$F$ give the magnification
and the phase shift, respectively.
In particular, in the high frequency limit (geometrical optics limit),
the phase shift of $F$ becomes proportional to the frequency
with its coefficient being $2\pi$ times the Shapiro time delay $\Delta t_g$.
We then define a quantity ${\hat \eta}$ by subtracting the Shapiro time delay from the phase shift, namely,
\be
{\hat \eta}(f) \equiv F(f)e^{-2\pi i f \Delta t_g}-1,
\ee
where the second term is added just to make ${\hat \eta}(f)$ vanish
in the absence of the lensing potential.
Observationally, it should be in principle possible to determine ${\hat \eta}$ for
the lensed GWs if the waveform covers a frequency
range including both the geometrical and wave optics regimes for which case we can set the phase
shift to zero in the high frequency side.
Continuous measurements of GWs from the evolving binaries may be 
a promising way for this purpose.

We introduce $K(f)$ and $S(f)$ by\footnote{To the first order in the gravitational potential, Eq.~(\ref{def-K-S}) leads to $K (f)=\operatorname{Re} (\eta (f))$ and $S(f)=\operatorname{Im}( \eta (f))$. The variances of $K(f)$ and $S(f)$
given by Eqs.~(\ref{Kf}) and (\ref{Sf}) are derived under this first-order approximation.}
\be
{\hat \eta} (f)+1=(1+K(f))e^{iS(f)}. \label{def-K-S}
\ee
In \citet{Takahashi:2005ug}, $K(f)$ was called amplitude fluctuation and $S(f)$ was called
phase fluctuation.
In what follows, we adopt these terminology.

Since the matter inhomogeneities are randomly distributed, 
$K(f)$ and $S(f)$ also behave as stochastic variables for each GW event.
Therefore, the statistical properties of these quantities are more useful to 
extract cosmological information than looking at the individual lensed events.
The two-point correlation function of $K(f)$ and $S(f)$ is related to the
power spectrum of $\Phi$ as \footnote{In \citet{Takahashi:2005ug, Oguri:2020ldf},
$\langle K^2 (f) \rangle$ and $\langle S^2(f) \rangle$ are expressed in terms
of the power spectrum of the matter density contrast $\delta$.
This can be done by going through the Poisson equation.
In Eqs.~(\ref{Kf}) and (\ref{Sf}), we instead use $P_\Phi$ since it is the metric perturbations
that cause the gravitational lensing and those equations (\ref{Kf}, \ref{Sf}) manifest that
they are free from the relation between the density contrast and the metric perturbations.
Please refer to the discussions in the next subsection regarding this point.
}
\begin{align}
&\langle K^2 (f) \rangle ={(4\pi f)}^2 
\int_0^{\chi_s} d\chi \int \frac{d^2 k_\perp}{{(2\pi)}^2} ~P_{\Phi}( k_\perp;t(\chi))
\sin^2 \left( \frac{r_F^2 k_\perp^2}{2} \right), \label{Kf} \\
&\langle S^2 (f) \rangle ={(4\pi f)}^2 
\int_0^{\chi_s} d\chi \int \frac{d^2 k_\perp}{{(2\pi)}^2} ~P_{\Phi}( k_\perp;t(\chi))
{\bigg[ 1- \cos \left( \frac{r_F^2 k_\perp^2}{2} \right) \bigg]}^2. \label{Sf}
\end{align}
Here $P_\Phi$ is the power spectrum of ${\tilde \Phi}$ defined by
\be
\label{P-phi}
\langle {\tilde \Phi} ({\bm k},t) {\tilde \Phi} ({\bm k'},t) \rangle
={(2\pi)}^3 P_\Phi (k,t) \delta ({\bm k}+{\bm k'}),
\ee
and $r_{\rm F}$ is the Fresnel scale \citep{Macquart:2004sh, Takahashi:2005ug}
\be
r_F^2 \equiv \frac{\chi (\chi_s-\chi)}{2\pi f \chi_s}.
\ee
The Fresnel scale is determined by the ratio of the geometrical distance
(\ref{geometrical-t}) to the GW wavelength.
This means the Fresnel scale provides a rough indication of the length scale
of the matter inhomogeneities below which
GWs are not sensitive to.
In mathematical language, this can be understood as the trigonometric functions appearing in the integration for the expressions
of $\langle K^2 (f) \rangle$ and $\langle S^2(f) \rangle$ 
and acting as erasing the contributions
of the modes shorter than the Fresnel scale.
Thus, frequency dependence of $\langle K^2 (f) \rangle$
and $\langle S^2(f) \rangle$ contains information of the matter power spectrum
at the Fresnel scale, and this fact enables us to probe the matter inhomogeneities on the Fresnel scale.
For the cosmological GW sources, a typical value of the Fresnel scale is 
\be
\label{Fresnel-2}
r_F \approx 130~{\rm pc}~{\left( \frac{f}{10^{-3} {\rm Hz}} \right)}^{-1}
\left( \frac{\chi(\chi_s-\chi)/\chi_s}{10~{\rm Gpc}} \right).
\ee
Thus, both space- and ground-based GW detectors can probe the matter inhomogeneities
on much smaller scales than on cosmological scales.
In particular, it has been suggested that the matter perturbations around the parsec scales
are dominated by dark low mass halos.
Measuring the amplitude and the phase fluctuations has the potential to probe how much
the dark matter is in the form of the gravitationally bound objects and shed light on
the nature of dark matter.

\subsection{Universal relation between the amplitude and the phase fluctuations}
In the previous subsection, we gave a brief overview of the
amplitude and the phase fluctuations of the GWs lensed by the
matter inhomogeneities.
Now, we point out that there is a consistency relation between the
variance of the amplitude fluctuation and 
that of the phase fluctuation as
\be
\langle K^2 (f) \rangle+\langle S^2 (f) \rangle
=\langle K^2 (2f) \rangle. \label{relation}
\ee
This relation can be straightforwardly derived from Eqs.~(\ref{Kf}) and (\ref{Sf}).
As far as we understand, this relation has not been given in the literature,
and Eq.~(\ref{relation}) is one of the main result of this Letter.
The purpose of this subsection is to discuss potential use and consequences 
of this relation.

Let us first clarify the fundamental assumptions we have made to arrive at the
consistency relation 
to figure out the domain in which the relation holds true and how universal the relation is.
The assumptions are as follows: 
(i) gravitational lensing is weak (i.e., Born approximation is valid), 
(ii) general relativity (GR) is valid, 
and (iii) statistical properties
of the matter perturbations respect the homogeneity and isotropy.
Let us consider these assumptions one by one.

First, as for the weak gravitational lensing, it has been demonstrated that the
typical amplitude of $K(f)$ and $S(f)$ is ${\cal O}(10^{-2})-{\cal O}(10^{-3})$ in a 
frequency range of our interests for the standard $\Lambda$CDM cosmology \citep{Oguri:2020ldf} \footnote{These typical magnitudes are values without accounting for selection of the lensing signal beyond a certain signal-to-noise ratio.}.
These amplitudes are much smaller than unity, and corrections from higher orders in the gravitational potential are suppressed more than the leading order one we have presented in the previous subsection.
Therefore, it is reasonable to expect that most lensing events are in the weak gravitational lensing regime.

Secondly, as for the point (ii),
the formalism presented in the previous subsection has been developed within
the framework of GR.
However, the consistency relation can hold true even for some alternative 
theories of gravity.
For instance, the simple class of the scalar-tensor theories only modifies the
scalar part of the metric perturbations, and the propagation equation for the
GWs is still the same as that in GR \citep{DeFelice:2011bh, Bellini:2014fua, Saltas:2014dha}.
In such a case, the background metric on which the GWs propagate can be written as
\be
\label{lens-metric}
ds^2=-(1+2\Phi)dt^2+a^2 (t) (1-2\gamma \Phi)d{\bm x}^2,
\ee
where $\gamma$ represents deviation from GR ($\gamma=1$ in GR if we ignore the anisotropic
stress of the matter).
In the scalar-tensor theories, $\gamma$ is in general a function of the cosmic time
and the length scales whose concrete form depends on the model under consideration \citep{Amendola:2007rr}.
Following the derivation of the amplification factor from the 
background metric given by Eq.~(\ref{lens-metric}) in the case of GR $(\gamma=1)$ \citep{Nakamura:1999} and applying it to the case with $\gamma \neq 1$,
we find that the modification to the amplification factor 
is only to replace $\Phi$ with $\frac{1+\gamma}{2}\Phi$.
Thus, both the amplitude fluctuation and the phase fluctuation are rescaled 
by the same factor, and the consistency relation is not modified.
Generally, modifying the scalar sector changes a relation between the
matter density contrast and the gravitational potential (i.e. Poisson equation: \citet{Amendola:2007rr})
and leads to the deviation of the growth rate of the matter inhomogeneities from GR,
but change of the matter power spectrum due to the different growth rate 
does not affect at all the expressions of $\langle K^2 (f) \rangle$ and $\langle S^2(f) \rangle$
since it is the gravitational potential that causes the gravitational lensing.
In this sense, not all the elements in GR are equally crucial to derive the relation (\ref{relation}).

Thirdly, the statistical homogeneity and isotropy, 
namely Eq.~(\ref{P-phi}), has been employed
to arrive at Eqs.~(\ref{Kf}) and (\ref{Sf}).
Violation of this assumption will in general lead to the violation of
the relation (\ref{relation}) but depending on how we abandon the assumption the 
consistency relation may still be satisfied.
For instance, dropping the statistical isotropy only replaces $P_\Phi (k_\perp;t(\chi))$
appearing in the integrands of Eqs.~(\ref{Kf}) and (\ref{Sf}) with the
direction-dependent one $P_\Phi ({\bm k_\perp};t(\chi))$.
We can easily verify that such a replacement does not change 
the consistency relation.

To summarize, there are a few assumptions imposed to derive the relation (\ref{relation}),
but depending on how one violates those assumptions the consistency relation can be still satisfied
under less restrictive conditions.
It is also worth mentioning that as it is clear from how the consistency 
relation has been derived,
it holds irrespective of the shape of the matter power spectrum.
Thus, the consistency relation does not rely on the physics that underlies the 
shape of the matter power spectrum such as the nature of dark matter 
and the early-universe models (inflation) characterizing the primordial power spectrum.

Having explained the underlying assumptions to obtain the consistency relation,
let us next discuss its potential applications.
One practical application would be to use the relation as an independent confirmation
that the observational determination of the variances of $K(f)$ and $S(f)$ 
from the measurements of many lensing events has been done correctly.
Observationally, measurements of the amplification factor is conducted by comparing 
the data with the unlensed template waveform.
Thus, extracting the amplification factor correctly can be achieved successfully only when
we have the correct theoretical modeling 
of the unlensed GW waveform that requires correct quantitative understanding of the GW sources.
In other words, incorrect modeling of the sources generates spurious contributions to $K(f)$ and $S(f)$.
Such undesired bias will lead to the violation of the consistency relation.
Given that the expected amplitudes of $K(f)$ and $S(f)$ are ${\cal O}(10^{-3})-{\cal O}(10^{-2})$ \citep{Oguri:2020ldf},
we naively suppose that the understanding of the GW sources at the same level is requisite
to avoid such bias.
Conversely, confirmation that the observationally determined 
$\langle K^2 (f) \rangle$ and $\langle S^2(f) \rangle$ satisfy the consistency relation 
provides us with a solid confidence that the information of the gravitational lensing has 
been obtained correctly, and we can safely use this observational result to
determine/constrain the small-scale power spectrum of the matter inhomogeneities.
In this sense, the consistency relation will be useful to establish the matter power spectrum
by the measurements of the lensing of GWs.

A second potential application of the consistency relation is to infer $\langle K^2 (f) \rangle$
at a higher frequency range that is out of the GW measurements.
The relation tells us that knowledge of $\langle K^2 \rangle$ and $\langle S^2 \rangle$
at a frequency $f$ enables us to infer $\langle K^2 \rangle$ at twice the frequency ($2f$).
Thus, it is possible to observationally determine $\langle K^2 \rangle$ up to twice 
the maximum frequency that GW detectors can reach.

A third potential application is a test of GR.
As we have already discussed, the consistency relation can be violated in alternative 
theories of gravity
for which the propagation equation of the GWs deviates from that in GR.
Therefore, observational verification of the relation provides a new test if the propagation
of GWs obey what GR predicts.
However, propagation of GWs has been already constrained to be very close to GR 
by the almost simultaneous detections of the GWs and gamma rays
from the neutron-star mergers \citep{TheLIGOScientific:2017qsa, Monitor:2017mdv},
and it is not clear how significantly the relation can in principle be 
violated by changing the 
propagation properties of the GWs without conflicting the existing constraints.
For the implications to the theories of modified gravity deduced from the recent GW observations, see, e.g. \citet{Creminelli:2017sry, Langlois:2017dyl}. 
After all, it is possible that the new test is not as strong as the other ones.
Even in that case, the consistency relation may be used as an independent confirmation of GR.
Clarifying this issue quantitatively is beyond the scope of this Letter.
To summarize this subsection, the universal relation between the amplitude and the phase
fluctuations has some interesting applications to help our understanding of the universe.

Before closing this subsection, there is one comment that may be worth noting.
The consistency relation is given in a simple form and fairly universal.
Thus, we expect that there is a clear physical explanation behind it, although
we were not able to find it.

\subsection{On the dependence of the amplitude and the phase fluctuations on the source distance}
Although not explicitly written (just for the sake of notational simplicity),
both $\langle K^2 (f) \rangle$ and $\langle S^2 (f) \rangle$ are functions of
not only $f$ but also the (comoving) distance to the source $\chi_s$.
In other words, the ensemble average $\langle \cdots \rangle$ is performed for fixed $f$ and $\chi_s$.
Observationally, the ensemble average is determined by measuring a sufficient number
of the GW events.
However, since each GW event has a different distance from us, 
strictly speaking, it is not possible to accumulate the GW events having exactly
the same distance.
Furthermore, there will be measurement errors of the distance for each GW event.  
As a compromise, we need to group the GW events having nearly the same source redshifts
to compute the ensemble average for a particular source redshift $z_s$.
This procedure may induce the error of the observational estimation of 
$\langle K^2 (f,z_s) \rangle$ and $\langle S^2 (f,z_s) \rangle$ (In this subsection, we explicitly show the dependence of $\langle K^2 (f) \rangle$
and $\langle S^2(f) \rangle$ on $\chi_s$ in terms of the corresponding source redshift $z_s$.).
According to the results presented in \citet{Takahashi:2005ug},
$\langle K^2 (f,z_s) \rangle$ and $\langle S^2 (f,z_s) \rangle$ change by ${\cal O}(1)$
by changing $z_s$ by ${\cal O}(1)$.
Thus, crudely speaking, the relative change of $\langle K^2 (f,z_s) \rangle$ and $\langle S^2 (f,z_s) \rangle$ by shifting the source redshift by $\Delta z_s$ 
is ${\cal O}(\Delta z_s)$.
However, since the consistency relation holds true at any $z_s$, we may circumvent the above
difficulty by considering the consistency relation integrated over some redshift range 
(e.g., $z_1 < z_s < z_2$) covering
the scatter of the redshifts of the GW events.
In other words, an estimator ${\cal E}$ constructed out of data $\{ K_i (f,z_{s,i}), S_i (f,z_{s,i}) \}_{i=1}^N$ of $N$ GW lensing events lying in the redshift range $(z_1, z_2)$ by
\begin{equation}
{\cal E}\equiv \frac{1}{N} \sum_{i=1}^N X_i,~~~~~~~~~~{\rm where}~~ X_i = K_i^2 (f,z_{s,i})+S_i^2 (f,z_{s,i}) - K_i^2 (2f,z_{s,i}),
\end{equation}
yields, if the consistency relation holds, $\langle {\cal E} \rangle =0$ and $\langle {\cal E}^2 \rangle \propto 1/N$ (Furthermore, for $N \gg 1$,
${\cal E}$ obeys the Gaussian distribution thanks to the central limit theorem.).
Therefore, if we alternatively consider the averaged consistency relation over some redshift range,
the accuracy of its observational confirmation is limited solely by $N$, i.e. the number of the
GW lensing events.

\begin{figure}[t]
  \begin{center}
    \includegraphics[clip,width=10.0cm]{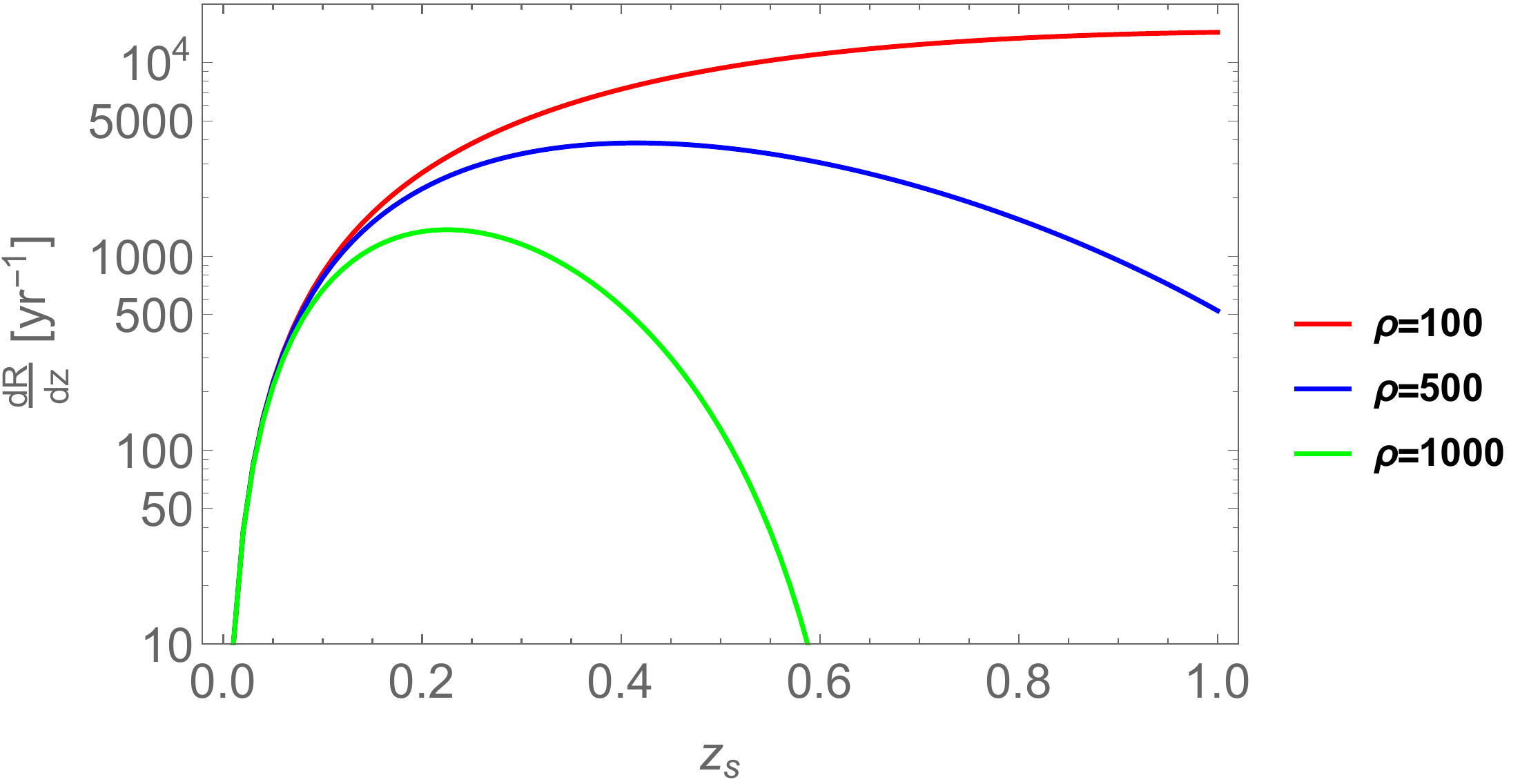}
    \caption{Differential merger rate detectable by DECIGO above 
    the signal-to-noise ratio $\rho=100, 500, 1000$.}
    \label{fig2}
  \end{center}
\end{figure}

In the above discussion, we have assumed that $K$ and $S$ for each GW lensing event are 
measured perfectly without errors.
In reality, both $K$ and $S$ are determined with some errors caused by the instrumental noises, 
which is another crucial
factor that hinders the observational verification of the consistency relation.
Detectability of the amplitude and the phase fluctuations taking into account the noises is partially discussed in \citet{Oguri:2020ldf}, but more detailed investigations remain to be clarified.
Since the main results in this work are to show the existence of the consistency relation 
and to propose its potential applications, we just give a crude estimation of $N$ for DECIGO
above a certain signal-to-noise ratio (SNR) $\rho$ following
the method presented in \citet{Ding:2015uha, Hou:2020mpr} before closing this subsection.
Fig.~\ref{fig2} shows the differential merger rate $\frac{dR}{dz}$ in the unit 
of ${\rm yr}^{-1}$ for which $\int_0^z \frac{dR}{dz} dz$ gives the detectable number of 
the merger events of equal mass binary with a chirp mass $26.5~M_\odot$ within the redshift $z$ above SNR $\rho=100, 500, 1000$.
Given that the GW amplitude can be measured with the accuracy $\sim 1/\rho$,
$\rho=1000$ will be a representative value for determining $\langle K^2 (f,z_s) \rangle$ and $\langle S^2 (f,z_s) \rangle$ whose typical values are about $10^{-3}$.
As an example, from Fig.~\ref{fig2}, we expect that one year of operation of DECIGO will detect
$\sim 100$ GW events having $\rho > 1000$ in the range $0.2 < z_s < 0.3$
(Since the relative error of the determination of the source distance is expected to be at most $0.01$ \citep{Camera:2013xfa}, the error of $z_s$ is not problematic if the redshift range
is taken to be larger.).
This suggests that with DECIGO we will be able to confirm the consistency relation
or detect its violation at ${\cal O}(1)$ level in terms of the relative difference.

\section{Conclusion}
Detections of the gravitational lensing of GWs are 
promising in the near future.
GWs from the cosmological distant sources are gravitationally lensed
by traveling through the matter inhomogeneities before arriving at detectors.
It has been proposed in the literature \citep{Takahashi:2005ug} that the variance of the modulation of the GW waveform
can be used to extract information of the matter power spectrum at very small scales
and the low-mass dark halos.
In this Letter, we have found that the variance of the amplitude fluctuation and
that of the phase fluctuation of the amplification factor obey a consistency relation
given by Eq.~(\ref{relation}).
This relation is universal in the sense that it does not rely on the shape of the
matter power spectrum.
We then investigated how universal the relation is and in which cases 
the relation can be violated.
We also discussed some potential applications of the consistency relation that
include the confirmation of the observational determination of the variances of the
gravitational lensing.
After the variances have been determined observationally over some frequency range,
the consistency relation may be useful to confirm the robustness
of their determinations and enables us to probe the matter spectrum at small scales with confidence 
by solving either Eq.(\ref{Kf}) or Eq.~(\ref{Sf}).

\section*{Acknowledgements}
We would like to thank Saul Hurwitz, Masamune Oguri, and Ryuichi Takahashi
for helpful comments.
This work is supported by the MEXT Grant-in-Aid for Scientific Research on Innovative Areas 
No.~17H06359~(T.S.) and No.~19K03864~(T.S.).

\bibliography{ref}

\end{document}